\documentclass[twocolumn,amsmath,amssymb]{revtex4}

\usepackage{graphicx}
\usepackage{dcolumn}
\usepackage{bm}
\usepackage{epstopdf}
\usepackage{epsfig}
\usepackage{multirow}
\usepackage[final]{pdfpages}

\begin{document}

\title{Genetic Design of Enhanced Valley Splitting towards a Spin Qubit in Silicon}

\author{Lijun Zhang}
\email{ljzhang13@gmail.com}
\affiliation{National Renewable Energy Laboratory, Golden, Colorado, USA 80401}
\author{Jun-Wei Luo}
\affiliation{National Renewable Energy Laboratory, Golden, Colorado, USA 80401}
\author{A. L. Saraiva}
\affiliation{Instituto de Fisica, Universidade Federal do Rio de Janeiro, Caixa Postal 68528, 21941-972 Rio de Janeiro, Brazil}
\author{Belita Koiller}
\email{bkoiller@gmail.com}
\affiliation{Instituto de Fisica, Universidade Federal do Rio de Janeiro, Caixa Postal 68528, 21941-972 Rio de Janeiro, Brazil}
\author{Alex Zunger}
\email{alex.zunger@gmail.com}
\affiliation{University of Colorado, Boulder, Colorado 80309, USA}

\date{\today}

\maketitle

{\bf Electronic spins in Silicon (Si) are rising contenders for qubits --
the logical unit of quantum computation~\cite{nielsen_quantum_2000} --
owing to its outstanding spin coherence properties and compatibility to standard
electronics~\cite{morton_embracing_2011, zwanenburg_silicon_2012}.
A remarkable limitation for spin quantum computing in Si hosts is the orbital degeneracy of this material's conduction band, preventing the spin-$\frac{1}{2}$ states from being an isolated two-level system~\cite{koiller_exchange_2001,zwanenburg_silicon_2012}.
So far available samples of Si quantum wells cladded by Ge-Si alloy barriers provide relatively small valley splitting (VS), with the order of 1 meV or less~\cite{nicholas_investigation_1980, lai_valley_2006, goswami_controllable_2007, boykin_valley_2004, friesen_valley_2007, saraiva_physical_2009}, degrading the fidelity of qubits encoded in spin ``up'' and ``down'' states in Si.
Here, based on an atomically resolved pseudopotential theory, we demonstrate that ordered Ge-Si layered barriers confining a Si slab can be harnessed to enhance the VS in the active Si region by up to one order of magnitude compared to the random alloy barriers adopted so far. 
A biologically inspired genetic-algorithm search~\cite{franceschetti_the_1999, piquini_band_2008} is employed to identify magic Ge/Si layer sequences of the superlattice barriers that isolate the electron ground state in a single valley composition with VS as large as $\sim$9 meV.
The enhanced VS is preserved with the reasonable inter-layer mixing between different species,
and is interestingly ``protected'' even if some larger mixing occurs.
Implementation of the optimized layer sequences of barriers, within reach of modern superlattice growth techniques~\cite{menczigar_enhanced_1993}, overcomes in a practical systematic way the main current limitations related to the orbital degeneracy, thus providing a roadmap for reliable spin-only quantum computing in Si.
}

\begin{figure}[ht!]
\resizebox{80mm}{!}{\includegraphics{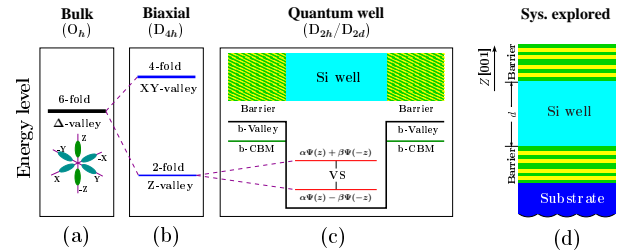}}
\caption{\textbf{Description of valley splitting and system explored.}
Schematic representation of (a) the (6-fold degenerate) $\Delta$-valley state of bulk Si; (b) splitting between $XY$-valley and $Z$-valley by tensile biaxial strain; and (c) splitting of $Z$-valleys by the abrupt interfaces of quantum well. Note that the confinement barrier height for the well is the band offset (between Si well and barrier) of $Z$-valley (b-Valley), higher than the band offset of conduction band minima (b-CBM). The symmetry of Si quantum well of $N$ monolayers alternates from D$_{2d} \leftrightarrow$ D$_{2h}$ for $N$ odd $\leftrightarrow$ even respectively.
(d) Heterostructure geometry adopted in the present study. A [001]-oriented Si quantum well of thickness $d$ and the surrounding barriers are coherently strained by epitaxial growth over a specified substrate. Both barrier and substrate are composed of Si-Ge based materials.}
\label{intro}
\end{figure}

The qubits for quantum information processing are encoded in two-level quantum systems {$\{|0\rangle,|1\rangle\}$},
and can be realized, for example,
by two spin states {$\{|\uparrow\rangle, |\downarrow\rangle\}$} of an electron at the conduction band edge of a semiconductor~\cite{loss_quantum_1998,kane_silicon_1998}.
While Si enjoys a number of advantages over III-V semiconductors in this respect,
including long spin coherence lifetime
(associated with its weak spin-orbit coupling and small content of nonzero-nuclear-spin isotopes),
as well as advanced fabrication know-how,
its major drawback is the (six-fold) orbital degeneracy of its lowest conduction band (Fig. \ref{intro}a)
located close to the $X$ point in the Brillouin zone.
This is no longer a two-level system determined solely by its spin,
leading to considerable leakage and
decoherence driven by the energetic proximity among the degenerate orbitals~\cite{koiller_exchange_2001}.
Whereas this six-fold valley degeneracy in the O$_h$-symmetric bulk Si
can be partially removed by application of tensile biaxial strain~\cite{schaffler_high_1997},
thus, isolating the two lowest $|+z\rangle$ and $|-z\rangle$ components from the rest (Fig. \ref{intro}b),
the creation of a sufficiently large energy splitting within this $Z$-valley subspace
(hereby called valley splitting (VS), see Fig. \ref{intro}c)
has proven to be a challenge for the experimental realization of Si-based spin qubits~\cite{zwanenburg_silicon_2012}.

The geometry of the basic physical system explored (Fig. \ref{intro}d)
includes a Si slab  (``Well'') interfaced by a material with higher conduction band (``barrier'').
The VS of this system depends on a multitude of degrees of freedom present in the
realistic device growth.
The Si well of thickness $d$ cladded by barrier materials of composition $X_b$ is coherently strained
on a substrate with the planar lattice constant $a_s$ (determined by its composition $X_s$).
We focus on the substrate and barrier composed of Ge-Si based materials,
which provide better-quality interfaces than oxides.
The barrier can be a Ge-Si random alloy of composition $X_b$
or any corresponding atomistically ordered structure.
We incorporate monolithically the full system containing up to 10$^{5}$ atoms/computational-cell,
via an atomistic pseudopotential Hamiltonian~\cite{zunger_first_1996, wang_local_1995},
solved in a plane-wave basis for each relaxed atomic configuration,
which gives directly the energies \{$\varepsilon_i$\} and
wavefunctions \{$\Psi_i$\} of the valley states.

\begin{figure}[ht!]
\resizebox{80mm}{!}{\includegraphics{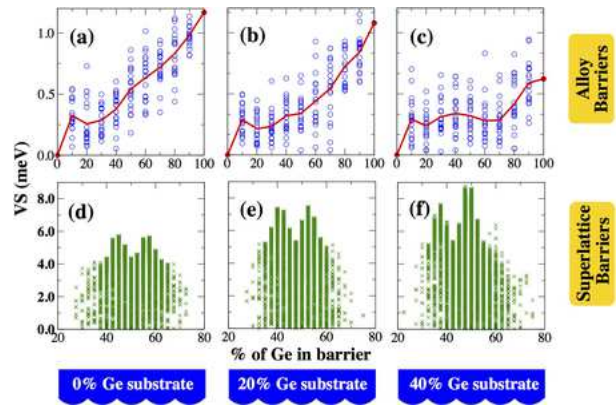}}
\caption{\textbf{Valley splitting of fixed-thickness Si well next to a barrier of disordered alloy and ordered superlattice.}
Calculated VS (in meV) as a function of the content of Ge in the barrier, for a 40 MLs Si quantum well embedded in Si-Ge disordered alloy (a,b,c) and ordered superlattice (d,e,f) barriers, on a 0$\%$, 20$\%$, and 40$\%$ Ge substrate.
Solid red lines in (a,b,c) represent the configuration-averaged VS for alloy barriers, and empty circles represent specific atomic realizations (20 for each composition).
In (d,e,f), the superlattice barrier has a 80-ML period, and structural configurations (green crosses) are generated by the genetic algorithm search (see Fig. \ref{ibs}). Note that the VS for superlattice barriers (d,e,f) is in $\sim$one-order larger scale than that of alloy barriers (a,b,c).}
\label{alloy_sl}
\end{figure}

\textbf{Macroscopic degrees of freedom:} We start by exploring the continuum
\emph{configuration-averaged} degrees of freedom in this system,
as common in the literature~\cite{boykin_valley_2004, nestoklon_spin_2006, friesen_valley_2007, valavanis_intervalley_2007, boykin_valley_2008, saraiva_physical_2009}, finding that whereas they do not provide a clear avenue to major VS enhancement,
their exploration hints at the importance of another length scale.
We consider a fixed-thickness Si well embedded in the Ge-Si alloy barriers with varied composition $X_b$,
on three substrates with different composition $X_s$.
For each alloy composition $X_b$ of barriers, we calculated 20 randomly realized atomic configurations and the averaged VS is evaluated.
The solid red line in Figs \ref{alloy_sl}a-c shows the calculated configuration-averaged VS as a function of composition $X_b$.
Generally, one observes an uneventful monotonic increase of the averaged VS
as the barrier becomes richer in Ge (see also Supplementary Fig. S1b, which shows the VS for a few distinct $X_b$).
Such continuum-like effect of the configuration-averaged alloy barriers
can be understood by the gradual change of the barrier height,
\emph{i.e.}, the band offset between the valley states of Si well and barrier
(``b-Valley'' in Fig. \ref{intro}c and Supplementary Fig. S1a)~\cite{saraiva_physical_2009, saraiva_intervalley_2011}.
Although the averaged VS (red lines in Figs \ref{alloy_sl}a-c) shows
substantial dependence on the epitaxial strain (also see Supplementary Fig. S1b),
the variation of the macroscopic barrier composition $X_b$ and substrate composition $X_s$ provides limited tuning of VS.

\textbf{Atomically resolved length scale:} Important clues emerge as to the significance of the atomically resolved length scale and symmetry,
as indicated in a recent work on the intervalley splittings of PbSe~\cite{poddubny_anomalous_2012}.
In principle, the splitting within the $Z$ valleys is closely related to the interface-induced
deviation from the O$_h$ symmetry of bulk Si (or the D$_{4h}$ symmetry of biaxially strained Si).
For a Si quantum well (Fig. \ref{intro}d),
the interfacial perturbation potential $\Delta V$ with the D$_{2h}$/D$_{2d}$ symmetry
provides a coupling channel between two $Z$-valley states,
giving a VS magnitude in perturbation theory of $2|\langle+z|\Delta V|-z\rangle|$.
To tune VS, we can engineer the magnitude and profile of the perturbation potential $\Delta V$
by varying the \emph{atomic-scale structure and symmetry} for the well and barriers.
The importance of the atomic scale is revealed, for example, in Fig. \ref{alloy_sl}a-c, where the blue circles represent
the VS obtained by resolving distinct random realizations of site occupations in alloy barriers.
The VS ranging from 0 to an upper bound of $\sim$1 meV is in reasonable agreement with experiments~\cite{nicholas_investigation_1980, lai_valley_2006, goswami_controllable_2007, zwanenburg_silicon_2012}.
We can see that the VS of Si can vary significantly
for different atomic configurations of barriers at the same composition $X_b$.
This is consistent with the recent calculation
showing that specific atomic arrangements at the interface region can result in
distinct VS (however the assumed Si$_{3}$Ge luzonite structure is difficult for experimental realization)~\cite{jiang_effects_2012}.
Also, the critical role of atomic resolution and symmetry
is apparent by considering a system of short-period Si-Ge superlattices located directly on a
substrate (i.e, no active Si layer in Fig. \ref{intro}d),
where our calculated VS reaches values as large as several tens of meV,
although the Si-Ge superlattice system is not the case of interest here (but
may relate to different qubit proposals~\cite{vrijen_electron_2000}).

Inspired by these basic insights from the atomic length scale,
we next explore in a systematic way whether and how atomic degrees of freedom
in the Si well, barrier composition and structure, and epitaxial substrate could raise the VS.
By varying the above degrees of freedom, we aim at identifying the rule of how the relevant physical factors govern VS,
and use it to seek an optimized VS.

\begin{figure}[ht!]
\resizebox{80mm}{!}{\includegraphics{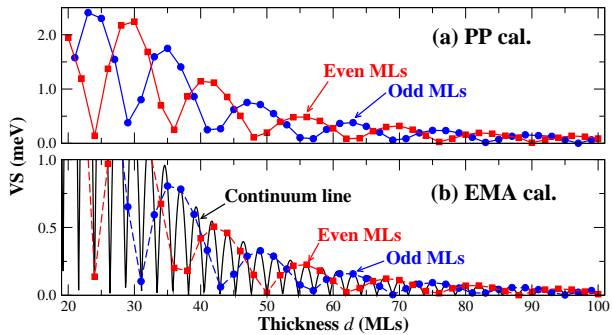}}
\caption{\textbf{Dependence of valley splitting on Si well thickness.}
Calculated VS (in meV) as a function of the thickness $d$ (in MLs) of a Si quantum well embedded in fixed pure Ge barrier on pure Si substrate. The pseudopotential (PP) and effective mass approach (EMA) results are shown in (a) and (b), and the data at the odd and even number of MLs are shown in blue and red, respectively. For comparison with PP data, we show the EMA results as a continuum line with markings at the integer multiples of the MLs spacing. Note that the EMA results are smaller in magnitude than the PP results.}
\label{o_e}
\end{figure}

\textbf{Effect of Si well thickness:} The thickness $d$ of the Si well is the first obvious parameter
to tune the perturbation potential $\Delta V$, and thus manipulate VS.
Fig. \ref{o_e}a shows the dependence of VS on the thickness $d$ in monolayers (MLs)
for fixed pure Ge barrier from the pseudopotential calculations.
We observe an overall decay in the magnitude of VS as the thickness $d$ increases,
while the VS for $d$ with an odd (blue circles) and even (red squares) number of MLs appears to
oscillate independently,
with a common period $\sim$14 ML and a phase shift of $\pi$/2.
This intriguing oscillatory behavior has been reported previously,
and was attributed to the symmetry change of the Si well
of $d$ MLs:  D$_{2d} \leftrightarrow$ D$_{2h}$ for $d$ odd $\leftrightarrow$ even~\cite{boykin_valley_2004, nestoklon_spin_2006, srinivasan_valley_2008}.

In Fig. \ref{o_e}b, we show the calculated VS within the effective mass approximation (EMA)
as a function of the continuum thickness $d_{con}$ (solid black line),
as well as the data sampled
at odd (blue circles) and even (red squares) atomic MLs.
We find that while the EMA results with continuum $d_{con}$ show a much faster oscillation,
clearly they reproduce well the existence of ``independent'' oscillation for discrete $d$ of odd and even MLs.
Thus, we attribute this atomic-scale odd-even independent oscillation
to a manifestation of the {\it aliasing effect}
(introduced by sampling a function at a rate
which is not fine enough to capture each oscillation),
rather than to a symmetry change (see Supplementary Information for detailed description).
This understanding underlines that to gain an optimized VS of Si well,
a well-controlled growth of monolayer precision
is required to reach the thickness $d$ at the peak of the oscillation.

\begin{figure}[ht!]
\resizebox{80mm}{!}{\includegraphics{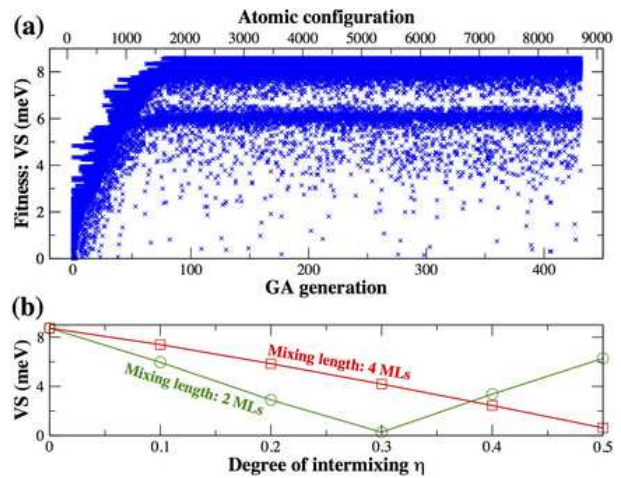}}
\caption{\textbf{Optimization of valley splitting in inverse-band-structure calculations and effect of inter-layer mixing.}
(a) Evolution of fitness (VS) with generation in an inverse-band-structure calculation for a 40 MLs Si quantum well embedded in ordered superlattice barriers with 80 ML period, on a 40$\%$ Ge substrate. The top axis shows the number of atomic configurations investigated during the evolution.
(b) Calculated VS (in meV) for the optimum configuration of superlattice barrier achieved in (a) -- Ge$_4$Si$_6$Ge$_2$Si$_6$Ge$_4$Si$_4$Ge$_4$Si$_4\ldots$, as a function of the degree of inter-layer mixing, $\eta$ (see text). At $\eta$ = 0 there is no mixing and $\eta$ = 0.5 means the maximum diffusion, i.e. complete destruction of Si-rich or Ge-rich pattern layer. We explore two cases of mixing length: 1ML for each side of the interface -- total of 2ML (green); and 2ML for each side -- total of 4ML (red), defining the maximum range at the interface where the mixing occurs. For each $\eta$, 10 alloy realizations are randomly sampled and the averaged VS is shown as a line. The fluctuation induced by different alloy realizations is within 0.2 meV.
}
\label{ibs}
\end{figure}

\textbf{Atomically ordered superlattices barriers:}
The substantial effect of specific atomic realization for the disordered alloy barriers (open circles in Figs \ref{alloy_sl}a-c)
stimulates us to investigate the situation
where the barriers are composed of ordered superlattices,
\textit{i.e.}, a repeated sequence of Si and Ge layers of arbitrarily assigned widths.
We explore the system composed of a 40 MLs Si well (located at an even peak of Fig. \ref{o_e}a)
embedded in the superlattice barriers with a period of 80 MLs
(with the minimum stacking unit of bilayer to comply with current experimental growth conditions).
This gives an astronomical number (2$^{40}$) of candidate layer-stacking configurations of barriers,
so a direct calculation for enumeration of all the candidates is not practical.
We perform an \emph{inverse-band-structure search} calculation~\cite{franceschetti_the_1999, piquini_band_2008}
where the best fitness is defined by the maximum VS,
and favorable structures are selected within a genetic algorithm approach.
Fig. \ref{ibs}a shows the evolution of fitness (VS) with generation (evolution step).
One clearly observes that the VS can be effectively tuned within a wide energy range,
from negligibly small up to $\sim$9 meV,
by varying Ge/Si stacking sequence of superlattice barriers.
Less than 100 generations already identify the best individuals,
which remain superior for the following hundreds of generations,
while new individuals still emerge with intermediate VS values.

Figs \ref{alloy_sl}d-f show the achieved VS of all the atomic configurations visited by the inverse-band-structure search,
sorted in terms of the Ge content in the barriers
on three varied substrates.
It is demonstrated that
a remarkable VS enhancement by a factor of 5-10 is achievable with ordered superlattice barriers as
compared to disordered alloy barriers (Figs \ref{alloy_sl}a-c) for all substrates.
Particularly, comparing with the maximum VS for the disordered alloy barriers
-- $\sim$1.0 meV on all the substrates,
the maximum VS (accompanied by the optimum configuration) for the ordered superlattice barriers reaches:
\centerline{$\%$ 0 Ge substrate: 5.7 meV (Ge$_4$Si$_4$Ge$_2$Si$_6$Ge$_4$Si$_4$Ge$_4$Si$_2\ldots$);\\}
\centerline{$\%$20 Ge substrate: 7.4 meV (Ge$_4$Si$_4$Ge$_4$Si$_2$Ge$_4$Si$_6$Ge$_4$Si$_2\ldots$);\\}
\centerline{$\%$40 Ge substrate: 8.7 meV (Ge$_4$Si$_6$Ge$_2$Si$_6$Ge$_4$Si$_4$Ge$_4$Si$_4\ldots$).\\}
We find that the multilayer superlattice barriers show larger VS around the central region,
\emph{i.e.}, at 40-60$\%$ Ge content in the barrier,
different from higher Ge content leading to larger VS for random alloy barriers.
The same Ge content in the superlattice barriers
can lead to both high and low VS extremes,
again emphasizing the key role of atomistic scale ordering in controlling VS.

\begin{figure}[ht!]
\resizebox{80mm}{!}{\includegraphics{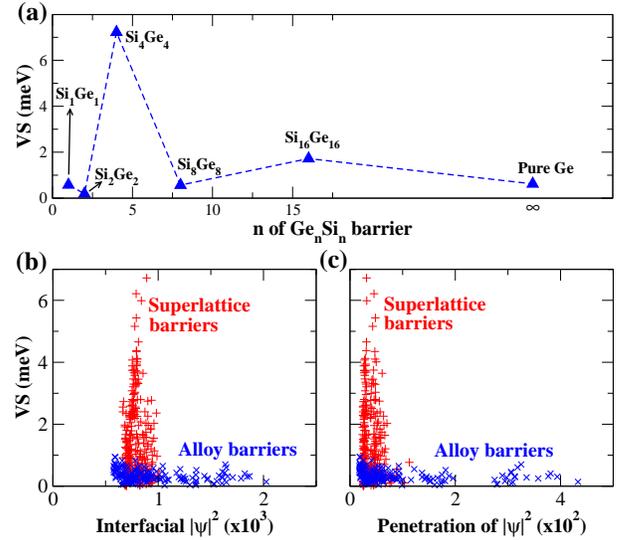}}
\caption{\textbf{Results of Ge$_{n}$Si$_{n}$ barriers, and relation between valley splitting and wave-function.}
(a) Calculated VS (in meV) for a 40 MLs Si quantum well embedded in the Ge$_{n}$Si$_{n}$ (n = 1, 2, 4, 8, 16) superlattice barriers on a 40$\%$ Ge substrate. The result for pure Ge barrier (n = $\infty$) is shown for comparison.
(b,c) Calculated VS (in meV) shown as a function of (b) planar-averaged (in the $XY$ plane) wave-function squared magnitude at the interface $z_I$ ($\overline{\Psi^2}(z_I)$), and (c) planar-averaged wave-function squared penetration into the barriers ($\int_{z_I}^\infty\overline{\Psi^2}(z) \mathrm{d}z$), for a 40 MLs Si quantum well embedded in random alloy and 16-ML period superlattice barriers, on a 40$\%$ Ge substrate. The averaged wave-function magnitude between two $Z$-valley states is used for such calculations.
}
\label{wav}
\end{figure}

\textbf{The Si/Ge$_4$ motif:}
Interestingly, all the optimum configurations identified start the barrier sequence by a Ge$_{4}$ sub-layer. This same``magic'' thickness for the first Ge sub-layer is also identified in the exhaustive enumeration calculations for the superlattice barriers with a shorter period of 16 MLs (see Supplementary Fig. S2a-c). Similar results are obtained for a Si well with the thickness of 47 MLs (located at an odd peak of Fig. \ref{o_e}a, see Supplementary Fig. S2d-f).
In order to better understand this, we explore a simple case -- the fixed 40 MLs Si well
embedded in Ge$_n$Si$_n$ superlattice barriers with $n$ = 1, 2, 4, 8, 16,
as shown in Fig. \ref{wav}a.
We see that the barrier of Ge$_4$Si$_4$ superlattice indeed exhibits the largest VS ($>$7 meV),
whereas all other barriers (including pure Ge) show typically low VS ($< 2$ meV).
This indicates that the starting sub-layer thinner or thicker than Ge$_4$
seem to equally suppress VS.
We unravel the underlying origin within the EMA context.
Briefly, the VS induced by a Si/Ge (ascending offset) interface has opposite sign to the Ge/Si
(descending offset) interface with the same wavefunction. Choosing the interface positions to match the
maxima/minima of the VS at the ascending/descending interfaces would maximize the total VS.  It is impossible
to match the interface positions perfectly to the incommensurate oscillations of well-thickness dependent VS (Fig. \ref{o_e}), but the Ge$_4$ sub-layer is the closest we can
get to this matching. Conversely, starting with a Ge$_2$ sub-layer cladding the Si well,
we find a destructive interference,
in agreement with the suppressed VS for Si$_2$Ge$_2$ superlattice barrier in Fig. \ref{wav}a.
This engineering is analogous to that of a \emph{distributed Bragg reflector} (see Supplementary Information for detailed description). But the fact that the oscillations
are incommensurate with the lattice and the strong dependence of VS on atomic ordering makes it impossible
to analytically predict the optimal structure. For this reason, the genetic selection of candidate structures is
an essential ingredient of this work.

Previous studies correlate the VS with the electronic wave-function magnitude
of the $Z$ valley at the interface (interfacial $|\Psi|^2$)~\cite{nestoklon_spin_2006, friesen_valley_2007, saraiva_physical_2009, saraiva_intervalley_2011}
and the wave-function penetration into the barrier region (penetration of $|\Psi|^2$)~\cite{saraiva_physical_2009, saraiva_intervalley_2011}.
In Fig. \ref{wav}b and \ref{wav}c,
we probed the VS as a function of interfacial $|\Psi|^2$ and penetration of $|\Psi|^2$,
for the 40 MLs Si well cladded by alloys barriers (blue crosses) and superlattice barriers (red pluses).
Compared with the alloy barriers,
the stronger confinement power of superlattice barriers
give much narrower distribution of both interfacial $|\Psi|^2$ and penetration of $|\Psi|^2$.
The optimum VS values for the superlattice barriers emerge
in the region of the narrowest distribution of these two quantities.
This is related to the sharp well/barrier interface for superlattice barriers,
which could in principle enhance the VS~\cite{saraiva_physical_2009,culcer_interface_2010}.

\textbf{Effect of Ge-Si intermixing in barriers:}
Since it is still a challenge to grow
perfectly pure sub-layer of Si or Ge in superlattices due to atomic inter-diffusion~\cite{menczigar_enhanced_1993},
we examine how much VS is affected by the \emph{interfacial mixing} between Si and Ge.
In particular, the inter-layer mixing is modeled by mapping pure Si into Si$_{1-\eta}$Ge$_\eta$,
and pure Ge into Ge$_{1-\eta}$Si$_\eta$ at the interfacial first few layers, determined by a mixing length.
The parameter $\eta$ quantifies the degree of inter-layer mixing, with $\eta$ = 0 corresponding to no mixing
and $\eta$ = 0.5 meaning maximum mixing,
\emph{i.e.}, complete destruction of Si-rich or Ge-rich pattern within this layer.
Fig. \ref{ibs}b shows the calculated VS as a function of $\eta$
for the above optimized superlattice barrier on $\%$40 Ge substrate
(Ge$_4$Si$_6$Ge$_2$Si$_6$Ge$_4$Si$_4$Ge$_4$Si$_4\ldots$, see Supplementary Fig. S3 for more ordered superlattice barriers), when
two cases of mixing lengths [2 MLs (green) and 4 MLs (red)] are explored.
Note that the favorable
Ge$_4$ starting sub-layer is only partially damaged if the mixing length is 2 ML (1 ML
towards each side of the interface), while for the 4ML mixing length the
Ge-pure layer is totally destroyed.
The non-trivial, non-monotonic behavior indicates that the intermixing
may lead to the formation of a more complex geometry which tunes VS by affecting
the interference pattern discussed before. This is reflected in a surprisingly steeper suppression of VS
in the shorter mixing length of 2 MLs compared to the longer mixing length of 4 MLs for small $\eta$.
Similarly, at very large
$\eta$, the structure becomes a complex layering of alloys, pure Si and pure Ge, which
might keep suppressing (the case of 4ML) or invert the symmetry and enhance the VS
(the case of 2ML). In both mixing lengths, for a reasonable degree of mixing ($\eta <$ 0.1),
the rather high VS of $>$6 meV is preserved.

\textbf{Advantage of atomically ordered barriers:}
We anticipate that the choice of ordered superlattice barriers instead of random alloy barriers might mitigate many
problems of real samples. For instance, the intrinsic non-deterministic nature of alloys induces disorder ranging
from the geometry of the interface plane to the inhomogeneous strain fields~\cite{evans_nanoscale_2012}.
The leakage of electrons tunneling through the superlattice barrier should also be reduced since the electronic density inside
the barrier is much reduced. Our atomically-resolved pseudopotential calculations
of the Si well strained on various substrates and interfaced with different barriers
can be used to explore the effects on VS of both global macroscopic quantities (strain, alloy compositions, geometric well thickness),
as well as atomic scale effects (ordering, inter-layer mixing, even-odd independent oscillation).
We identify the critical Si well thicknesses
as well as an emerging ``magic'' motif of $Ge_4$ starting in the ordered superlattice barrier
that causes strong coupling between $|+z\rangle$ and $|-z\rangle$ valley states,
leading to significantly enhanced VS as large as $\sim$9 meV.
The predicted structure is accessible within current experimental fabrication capabilities.
This opens the way to fundamental understanding of
the hitherto rather intangible $Z$-valley splitting in indirect-gap semiconductors such as Si
with the possible benefit of isolating single electron valley state for spin-based quantum computing.

\bigskip
\bigskip

\textbf{Methods}

The structures employed to optimize VS in this work (Fig. \ref{intro}d)
involve an active Si well with the thickness of $d$ MLs,
cladded on both sides by a barrier consisting of Si-Ge based materials,
including homogeneous random alloy and layer-by-layer superlattice structures.
The whole system is coherently strained on a substrate,
via minimization of atomically-resolved strain
with a generalized valence force field method (see Supplementary Methods A)~\cite{bernard_strain_1991}.
To comply with what is currently accessible in experimental growth we used two restrictions:
(i) Since too high Ge content in substrate is known to cause dislocations in thick Si active layers to relieve excessive strain,
up to 40$\%$ Ge content is considered in substrate.
(ii) a bilayer is used as the minimum stacking unit of each specie (Si/Ge) for the superlattice barrier.

The energies and wave-functions of conduction valley states for candidate structures
are calculated ``on the fly'' with the atomistic pseudopotential method,
described in detail in Refs.~\cite{wang_local_1995, zunger_first_1996}.
The atomistic pseudopotential method (overcoming the well-known Density-Functional-Theory limitations on electronic structure calculations),
accompanied with a plane-wave basis set and folded-spectrum diagonalization,
allow us to accurately calculate energy splitting of $Z$-valley states (at the order of meV or lower)
for numerous candidate structures with economic efficiency, as described in Supplementary Methods B.

Effective mass approach calculations were performed to accompany the
interpretation of the pseudopotential results. The effective mass calculations follow essentially the
model presented in Refs.~\cite{saraiva_physical_2009, saraiva_intervalley_2011}, adapted
to describe quantum wells in first order approximation, as described briefly
in Supplementary Methods C.

For Si wells embedded in layer-by-layer superlattice barriers,
since the search space shows a combinatorial burst of degrees of freedom,
we employ
the developed inverse-band-structure approach~\cite{franceschetti_the_1999, piquini_band_2008, davezac_genetic_2012},
\emph{i.e.}, a biologically inspired (Darwinian) genetic algorithm to guide the electronic structure calculations,
with the aim at finding the optimum configuration that gives the maximum VS (Supplementary Methods D).

\bigskip

\textbf{Acknowledgements}

We thank M. A. Eriksson and M d'Avezac for helpful discussions.
This work is supported by the U.S. Department of Energy,
Office of Science, Basic Energy Sciences, under Contract No. DE-AC36-08GO28308 to NREL.
The "Center for Inverse Design" is a DOE Energy Frontier Research Center.
A.S. and B.K.'s work is part of the
Brazilian National Institute for Science and Technology
on Quantum Information. A.S. and B.K. acknowledge
partial support from FAPERJ, CNPq and CAPES.

\bibliography{ref_vs}
\end{document}